\documentclass[12pt]{article}

\usepackage[utf8]{inputenc}
\usepackage[T1]{fontenc}
\usepackage{hyperref}
\usepackage{url}

\hypersetup{
  colorlinks=true,
  linkcolor=blue,
  citecolor=blue,
  urlcolor=blue
}

\title{\textbf{Learning Together: A Format for Reflective Turn-Based Sharing in STEM Communities}}

\author{
  James Day,$^{1}$ Katherine R.\ Herperger,$^{1}$ and Kyle Monkman$^{2}$ \\[6pt]
  \small $^{1}$Quantum Matter Institute, University of British Columbia, Vancouver, BC, Canada \\
  \small $^{2}$University of Calgary, Calgary, AB, Canada \\
}

\date{}

\begin{document}

\maketitle

\section*{Introduction}

Here, we present \textit{Learning Together}: a simple, low-cost format for structured speaking and
listening on historical, cultural, and equity-related topics within a physics community. In this
article, we describe the process of running these hour-long sessions, including what is needed,
how to set expectations for participants, and practical facilitation moves that help participants
reflect safely on unfamiliar and sometimes difficult material. We aim to offer a replicable recipe
that instructors and departments can adapt to their own contexts.

A critical component of implementing Equity, Diversity, and Inclusion (EDI) is understanding the
unique history and challenges pertaining to all members of a community. After all, physics is done
by people, in places, on land with histories. Physics communities have traditionally prized a
culture of objectivity that can make explicitly social and historical conversations feel out of
place, and have only recently begun to grapple with underrepresentation patterns that are among
the most acute in STEM. The logic behind a land acknowledgement which precedes an event can be
applied more broadly here: local history and current events are relevant to any physics community
situated in that place.\textsuperscript{1,2}
The structured speaking and listening format we describe draws on the tradition of talking
circles,\textsuperscript{3,4} a foundational approach to Indigenous pedagogy that centers
relational listening, reflective witnessing, and the co-creation of understanding. In Science,
Technology, Engineering, and Math (STEM) academic settings, there is an increasing body of
work\textsuperscript{5,6,7,8} that makes clear how inequities can create situations that lead to
lower retention and persistence. Many departments, however, still lack structured spaces for
intentional reflection on systemic issues and diverse worldviews. Without such spaces, EDI
policies may not be fully integrated into daily academic life, and exclusionary norms may
continue.\textsuperscript{9,10,11}
We recognize that in some climates, activities of this kind may feel risky or unsuitable for
facilitators and participants, and we encourage readers to assess their own context before adopting
this format.

Our initiative, entitled \textit{Learning Together}, creates space within a STEM community for
open, relational turn-based sharing on historical, cultural, and equity-related topics. For us,
this is the Quantum Matter Institute (QMI) at the University of British Columbia (UBC): a
university-based research center focused on quantum materials and condensed matter physics,
housing faculty, postdoctoral researchers, graduate students, and staff from multiple departments.
Unlike a single physics department, it draws members across disciplinary and administrative lines,
creating a community defined by shared research focus rather than a common degree program. This
paper outlines a step-by-step process for scientists and science-adjacent
individuals---students, researchers, instructors, and administrators---who recognize the ethical
necessity of doing EDI work but are unsure of how to begin in STEM disciplines.

\section*{What You Need}

Implementing \textit{Learning Together} requires minimal-to-no institutional funding; only an
institution-wide email, a room booking, and a session plan are needed. While we run these sessions
in a research institute, the idea can be implemented, for example, in a Teaching/Learning
Assistant Community of Practice, a department teaching meeting, or a high-school physics
Professional Learning Community. In our case, we facilitate four to six sessions per year (enough
to maintain momentum but not so much as to drain capacity). For us, facilitation is shared between
two colleagues who co-developed the format; we do not rotate facilitators, as consistency helps
build participant trust over time, but in other contexts, a rotating model might work fine.
Facilitators need not be experts in the session topic; familiarity with the format and a
commitment to the process are sufficient. The steps to create a \textit{Learning Together} session
are as follows:

\begin{enumerate}

  \item \textbf{Select a topic.} In our case, the facilitators selected topics that focused either
  on EDI challenges in STEM contexts or on discrimination issues in the local area.

  \item \textbf{Identify relevant short articles and/or videos.} The goal of these pre-session
  materials is not to be comprehensive but to ignite curiosity. Participants expect to spend 20 to
  30 minutes reading preparatory material before the session.

  \item \textbf{Send an invitation email.} An email is sent to the institute one week in advance,
  explaining the purpose of the session and sharing the pre-session material. All department
  members are encouraged to engage with the material, whether or not they plan to attend.
  Conversely, engagement with the pre-session material is not required for attendance.

  \item \textbf{Book a room.} We suggest finding a room in your institute with a circular seating
  arrangement.

  \item \textbf{Prepare slides.} For instance, a Land Acknowledgement slide and a small number of
  overview and prompt slides to guide the session. We recommend a minimal number of slides to save
  time and to emphasize that everyone is ``learning together,'' with no one teaching the material.

\end{enumerate}

Here, we describe how a \textit{Learning Together} session typically unfolds:

\begin{enumerate}

  \item \textbf{Welcome \& Context (10 min):} We share our slides, beginning with a land
  acknowledgement and a summary of the pre-session materials (serving as a reminder for those who
  engaged with them and an overview for those who did not).

  \item \textbf{The Prompt (5 min):} We finish our slides with a single prompt to anchor the
  session. The prompt is designed to encourage reflection and to share our personal understanding
  of these topics: ``What would you like to express about your experience with this
  content/resource? Feel free to respond to something you've heard, without offering advice or
  trying to solve/fix.'' We then emphasize that this is not a debate, nor is it even a
  conversation. A minute or two is taken in silence as participants organize their reactions.

  \item \textbf{Reflection (40 min):} Facilitators and participants take turns reflecting, starting
  with whoever feels ready. We are usually seated around a table in a circle and proceed clockwise.
  Participants are reminded that they do not need to be coherent when speaking and may pass on
  their turn. Depending on the number of people in the session, we may complete the circle once or
  twice.

  \item \textbf{Closing (5 min):} We thank the participants for learning with us and encourage
  them to continue learning and talking.

\end{enumerate}

\section*{Suggestions for Facilitation}

Facilitators should remind participants that empathy while listening is integral. Everyone is asked
to be charitable in their interpretations of what others say, as the speaker is likely formulating
their ideas in real time and might not yet have developed self-consistent thoughts on the matter.
Many of the words spoken may not yet reflect deep deliberation and may be very early-stage
reactions. Learning to sit with silence and some discomfort is important. Facilitators must also
uphold the convention that no interruptions or interjections be made. The current speaker's time
must be respected; they have been told not to worry about the coherence of their words and might
require multiple attempts to work through and consider their perspective. Everyone will eventually
get their turn to speak.

A reader adopting this format might be wondering: ``What if someone says something actively
harmful?'' It is worth noting that the people who choose to attend our sessions are also among the
least likely to speak recklessly. Still, while we have been fortunate not to have encountered this
scenario yet, we have given it some thought. While our ``no interjections'' norm is central, there
must exist an exception protocol so that harmful remarks do not stand in the moment. In the rare
case that a comment risks harm, facilitators could pause the circle and briefly restate the core
norms: empathetic listening for developing thoughts, no rebuttals, speaking from one's own
experience (avoiding claims about groups), and attention to impact over intent. This should
reassure participants that the format is safe and responsible.

There are a few issues we have encountered. The first concerns an expectation mismatch. Many
participants may initially expect a debate or lecture. This concern can be mitigated by clearly
articulating the norms governing each session: participants are under no obligation to speak,
rebuttals are excluded, and coherence is not required, as the primary objective is reflection.
Secondly, these sessions can center around heavy topics. Discussions of human rights and systemic
inequality, for example, carry significant weight. We suggest reducing overall sensitivity by
facilitating with trusted colleagues and occasionally mixing lighter articles with heavier ones.
Thirdly, scheduling conflicts and a busy institute calendar can affect attendance. A solution can
involve intentional timing, perhaps by inviting those who set the tone and culture to attend and
coordinating with their schedules. Their presence signals that this work is institutionally valued;
the meaningful middle ground is the busy colleague who deprioritizes these sessions without
opposing them, and their occasional attendance carries weight. Similarly, a final issue we noticed
concerns minimal attendance in some sessions. For us facilitators, it is important to remember
that some people will still interface with the pre-session materials even if they do not attend the
session (we know this anecdotally). Sharing the invitation with the entire institute guarantees
that everyone has access to the resources, even if they do not attend the session. Simply raising
awareness amongst an institute that the given issue exists is better than doing nothing. It is also
beneficial to emphasize that people who do not personally identify with the demographic or history
being discussed can still gain valuable insight and contribute meaningfully to the gathering.

\section*{Evidence of Value}

Conversations with some of our participants outside the \textit{Learning Together} sessions, along
with written reflections from our facilitators, have convinced us that this effort is worthwhile.
Several attendees linked the themes to their own histories, recognizing personal or familial
connections to the topics and creating meaningful moments of resonance. For another participant,
witnessing allyship mattered: ``It meant a lot to me to see non-Indigenous people think
meaningfully about Indigenous perspectives.'' One newcomer to equity work emphasized design over
debate: ``The structured, non-confrontational format created a safer, more accessible space for
those new to these conversations.''

For those who try to implement our format, remember to keep focus on the overall goal, despite
potential setbacks. Low attendance and limited progress do not mean that nothing good is happening.
This work is a journey, not a destination; the more we learn together, the longer we see the
journey will be.

This type of work is challenging and ongoing, requiring stamina, tolerance to discomfort, and a
long-term commitment. We have created space for people to feel and stay with discomfort. Repeated
attendance has helped to build stamina, counter isolation, and perhaps even foster some
accountability. The goal is not to execute this process perfectly, but to remain present and
engaged when the work becomes difficult.

\section*{Our Topic List}

Topics to date have covered the histories and legacies of settler colonialism, systemic racism,
and community resilience in British Columbia. Specifically:

\begin{itemize}
  \item The relationship between UBC and the Musqueam First Nation
  \item The experiences of newcomers and immigrants to Vancouver
  \item The Chinese Head Tax and Exclusion Act
  \item The internment of Japanese-Canadians during WWII
  \item The Komagata Maru incident and South Asian community history in Canada
  \item The legacy of Hogan's Alley
  \item The ongoing challenges facing women in science
  \item 2SLGBTQIA+ history of Vancouver
  \item Settler--Indigenous relations: learning, commemoration, and action
  \item Anti-Asian racism across North America, exacerbated by the COVID-19 pandemic
  \item The opioid crisis in Vancouver
\end{itemize}

\section*{Conclusion}

In general, the \textit{Learning Together} format is designed to create a place to discuss
inequities in science and diverse worldviews. While this format has been run in a research
institute, the underlying structure (a circle, a prompt, turn-based reflection) has been applied
across a range of educational and community contexts,\textsuperscript{3} and may be adaptable to
other STEM settings (a graduate student association, a lab group meeting, a teaching community of
practice, etc.), though facilitators should attend carefully to power dynamics when participants
occupy different institutional roles. This initiative is an example of institutional progress that
can be made with minimal funding and resources.

\section*{Author Bios}

\noindent\textbf{James Day} (he/him) holds a PhD in Physics from the University of Alberta and is
a Research Associate at the Quantum Matter Institute at the University of British Columbia, where
he works in experimental condensed matter physics and leads science outreach and EDI initiatives.
As someone working in a field with well-documented representation gaps, he sees structured EDI
dialogue as both a professional responsibility and a personal commitment.

\medskip
\noindent\textbf{Katherine R.\ Herperger} (she/her) earned her Honours BSc in Physics from the
University of Ottawa, where she conducted research in both computational chemistry and experimental
laser physics. She then moved to the University of British Columbia, completing her MSc in Physics
with a focus on quantum scattering, before beginning her PhD in Physics, where she now studies
diblock copolymer dynamics under the supervision of Dr.\ Joerg Rottler. Throughout each stage of
her academic career, she has remained deeply committed to advancing equity, diversity, and
inclusion in academia.

\medskip
\noindent\textbf{Kyle Monkman} (he/him) is a Red River M\'{e}tis and Ukrainian-Canadian physicist
who was born and raised in the Red River Valley of Winnipeg, Manitoba. He was introduced to
physics through the Engineering Access Program (ENGAP) for Indigenous students in engineering at
the University of Manitoba. He has since completed his PhD in Physics at the University of
Manitoba and is now a postdoctoral researcher at the University of Calgary.

\section*{Notes and References}

\begin{enumerate}

  \item S.~Stein, C.~Ahenakew, W.~Valley, P.~Y.~Sherpa, E.~Crowson, T.~Robin, W.~Mendes, and
  S.~Evans, ``Toward more ethical engagements between Western and Indigenous sciences,'' FACETS
  \textbf{9}, 1--14 (2024).

  \item J.~Seniuk~Cicek, R.~Herrmann, R.~Forrest, and K.~Monkman, ``Decolonizing and
  Indigenizing Engineering: The Design \& Implementation of a New Course,'' Proc.\ Can.\ Eng.\
  Ed.\ Assoc.\ (2022).

  \item P.~Barkaskas and D.~Gladwin, ``Pedagogical Talking Circles: Decolonizing Education
  through Relational Indigenous Frameworks,'' J.\ Teach.\ Learn.\ \textbf{15}(1), 20--38 (2021).

  \item ``Circle Pedagogies: Indigenous Pedagogy in the classroom,''
  \url{https://uwaterloo.ca/centre-for-teaching-excellence/catalogs/integrative-and-experiential-learning-series/circle-pedagogies-indigenous-pedagogy-classroom}.

  \item E.~Corrigan, M.~Williams, and M.~A.~Wells, ``Beyond gender: The intersectional impact of
  community demographics on the continuation rates of male and female students into high school
  physics,'' Can.\ J.\ Phys.\ \textbf{102}, 577--589 (2024).

  \item American Institute of Physics (TEAM-UP Task Force), \textit{The Time Is Now: Systemic
  Changes to Increase African Americans with Bachelor's Degrees in Physics and Astronomy} (AIP,
  College Park, MD, 2020).

  \item ``Effective Practices for Physics Programs (EP3): Equity, Diversity, and Inclusion,''
  \url{https://ep3guide.org/guide/equity-diversity-and-inclusion/}.

  \item E.~Hennessey, A.~Smolina, S.~Hennessey, A.~Tassone, A.~Jay, S.~Ghose, and K.~Hewitt,
  ``Canadian Physics Counts: An exploration of the diverse identities of physics students and
  professionals in Canada,'' FACETS \textbf{10}, 1--16 (2025).

  \item American Physical Society, \textit{LGBT Climate in Physics: Building an Inclusive
  Community} (APS, College Park, MD, 2016).

  \item R.~S.~Barthelemy, M.~McCormick, and C.~Henderson, ``Gender discrimination in physics and
  astronomy: Graduate student experiences of sexism and gender microaggressions,'' Phys.\ Rev.\
  Phys.\ Educ.\ Res.\ \textbf{12}, 020119 (2016).

  \item A.~M.~Hallas, ``Underrepresentation of women last authors in Nature Physics,'' Nat.\
  Phys.\ \textbf{21}, 491--493 (2025).

\end{enumerate}

\end{document}